\documentclass[journal]{IEEEtran}
\usepackage{amssymb,amsmath,epsfig,graphics,graphicx,subfigure,array}

\usepackage{epsfig}
\usepackage{bm}
\usepackage{url}
\usepackage{float}
\usepackage{subfigure}
\usepackage{graphicx}
\usepackage{capt-of}
\usepackage{lipsum}
\usepackage{tabularx}
\usepackage{array}
\usepackage{caption}
\usepackage{booktabs}
\usepackage[table]{xcolor} 
\newcolumntype{Y}{>{\centering\arraybackslash}p{2cm}} 
\usepackage{array}
\usepackage{threeparttable}

\usepackage[ruled,vlined,linesnumbered]{algorithm2e}
\DontPrintSemicolon
\SetKwRepeat{Do}{do}{while}%

\usepackage[noend]{algpseudocode}
\usepackage{multirow}
\usepackage{cleveref}

\usepackage{booktabs}
\usepackage[table]{xcolor}
\usepackage{array}
\newcolumntype{Y}{>{\centering\arraybackslash}p{2.5cm}}

\renewcommand{\arraystretch}{1.3}
\newcolumntype{Y}{>{\centering\arraybackslash}p{2cm}}

\usepackage{color}
\usepackage{amsmath}
\usepackage{tabularx}
\usepackage{epstopdf}
\usepackage{epsfig}

\newcommand{\warn}[1]{}

\makeatletter
\def\BState{\State\hskip-\ALG@thistlm}
\makeatother

\begin{document}

\title{Breaking Android with AI: A Deep Dive into LLM-Powered Exploitation}

\newcommand{\superast}{\raisebox{9pt}{$\ast$}}
\newcommand{\superdagger}{\raisebox{9pt}{$\dagger$}}
\newcommand{\superddagger}{\raisebox{9pt}{$\ddagger$}}
\newcommand{\superS}{\raisebox{9pt}{$\S$}}
\newcommand{\superP}{\raisebox{9pt}{$\P$}}

\author{\IEEEauthorblockN{Wanni Vidulige Ishan Perera\IEEEauthorrefmark{1}, Xing Liu\IEEEauthorrefmark{1},Fan liang\IEEEauthorrefmark{1}, Junyi Zhang}\\
\IEEEauthorblockA{\IEEEauthorrefmark{1}Sam Houston State University, USA \\
Emails: \{wdp006,xxl020u,fxl027\}@shsu.edu zjywy0228@gmail.com}}
\maketitle

\begin{abstract}
The rapid evolution of Artificial Intelligence (AI) and Large Language Models (LLMs) has opened up new opportunities in the area of cybersecurity, especially in the exploitation automation landscape and penetration testing. This study explores Android penetration testing automation using LLM-based tools, especially PentestGPT, to identify and execute rooting techniques. Through a comparison of the traditional manual rooting process and exploitation methods produced using AI, this study evaluates the efficacy, reliability, and scalability of automated penetration testing in achieving high-level privilege access on Android devices. With the use of an Android emulator (Genymotion) as the testbed, we fully execute both traditional and exploit-based rooting methods, automating the process using AI-generated scripts. Secondly, we create a web application by integrating OpenAI's API to facilitate automated script generation from LLM-processed responses. The research focuses on the effectiveness of AI-enabled exploitation by comparing automated and manual penetration testing protocols, by determining LLM weaknesses and strengths along the way. We also provide security suggestions of AI-enabled exploitation, including ethical factors and potential misuse. The findings exhibit that while LLMs can significantly streamline the workflow of exploitation, they need to be controlled by humans to ensure accuracy and ethical application. This study adds to the increasing body of literature on AI-powered cybersecurity and its effect on ethical hacking, security research, and mobile device security.
\end{abstract}


\IEEEpeerreviewmaketitle

\section{Introduction}\label{intro}
The growing utilization of mobile technology has significantly expanded the potential attack surfaces of cyber attacks, and Android devices are a high-priority target for exploitation. The open nature of the Android operating system, coupled with extensive usage, combines it highly vulnerable to a large number of attack surfaces, such as privilege escalation, bootloader attacks, and kernel exploits. Recent research discovered fundamental security weaknesses of the Android platform, such as the threat that native code vulnerabilities pose \cite{sanna2024risk} and the implementation defects in custom permission design that contribute to privilege escalation \cite{li2021android}. Further, research work on hardware security\cite{munoz2024cracking} has pointed out critical vulnerabilities that can be exploited to attack Android devices at the bootloader stage, so it makes worse day by day the security threats related to unauthorized root access.

Traditional penetration testing in the context of Android security is marked by its manual and time-consuming nature, generally requires a considerable technical expertise to effectively discover and exploit vulnerabilities. However, the latest developments in artificial intelligence, specifically in Large Language Models (LLMs), create new pathways for penetration testing automation. PentestGPT, introduced by Deng et al. \cite{deng2023pentestgpt}, has shown the possibility for LLMs to provide security assessment by automating vulnerability discovery and exploitation. Similarly, recent research aiming at AI-driven autonomous exploitation, such as in VulnBot \cite{kong2025vulnbot}, demonstrates the promise of multiagent cooperative systems to conduct penetration testing with little human effort. In addition, the use of generative AI for penetration testing \cite{hilario2024generative} has explained not only the advantages but also the ethical considerations associated with AI-assisted security testing and underscored the need to balance automation and responsible oversight.

With these developments, there are many significant challenges for the complete automation of penetration testing for Android platforms. Artificial intelligence-driven models currently continue to struggle with contextual analysis in complicated security environments with the need for human intervention to confirm and specify exploitation methods. Furthermore, ethical concerns in the application of AI for cybersecurity, particularly its misuse to create autonomous attack systems \cite{fang2024llm} entail careful implementation. LLM research on its own exploiting one-day vulnerabilities \cite{fang2024llm} has also generated concerns about the ability of AI to weaponize discovered vulnerabilities, again highlighting the need for security systems with ethical dimensions and robust control mechanisms.

This research aims to bridge these gaps by designing a systematic approach to the use of LLMs in Android penetration testing with respect to security and ethical issues. Specifically, we present a novel framework integrating LLM-based automation of rooting techniques, and privilege escalation. By performing an empirical analysis in an android emulator which is Genymotion, we assess the efficiency, accuracy, and security implications of LLM-assisted exploitation techniques.

The remainder of this paper is structured as follows: Section II provides a literature review of existing Android exploitation techniques and AI applications in penetration testing. In Section III we represent the methodology of the research which is automating rooting with the help of AI models. In Section IV we introduce the experimental setup and results, comparing the performance of LLM-based exploitation techniques. Section V discusses the ethical implications and security concerns in automated penetration testing. Lastly, Section VI summarizes the study with major findings and recommendations for future research.

\section{Related Works}\label{related}

The field of Android security has been studied extensively, and researchers have identified many vulnerabilities in privilege escalation, bootloader security, and kernel exploits. Li et al. \cite{li2021android} examined security vulnerabilities in Android custom permissions and found serious privilege escalation risks that attackers can exploit. Similarly, Meng et al. \cite{meng2019analyzing} experimentally examined how high-privilege Android apps, such as screenshot and screen recording apps, abuse their permissions to perform unauthorized actions. Their findings demonstrate the risks of unacceptable permission grants and how they would be used for privilege escalation attacks. 

Similarly, Sanna et al. \cite{sanna2024risk} conducted a risk estimation study of native code vulnerabilities in Android applications, describing how insecure memory operations in Android's native layers pose severe security risks. Muñoz \cite{munoz2024cracking} also looked into hardware vulnerabilities, describing how insecure configurations at the hardware level allow attackers to bypass Android's secure boot mechanisms.

Furthermore, beyond hardware and software vulnerabilities, existing literature examined approaches to analyze and emulate the Android boot process for security evaluation. A study conducted by Bertels et al. \cite{bertels2022emulating} showed how boot-emulation techniques can reveal security vulnerabilities in early Android firmware, hence offering valuable insight on how to secure the bootloader and reduce the risk of unauthorized modifications.

Recent findings by Happe et al. \cite{happe2023llms} have raised alarm regarding AI's potential to automatically exploit known vulnerabilities. Their work on LLM-based autonomous privilege escalation attacks illustrates how AI models can potentially generate and run exploits for Linux-based privilege escalation vulnerabilities and shows the ethical issues of AI-augmented cybersecurity operations.Happe and Cito's \cite{happe2023getting} study analyzed the use of generative AI for penetration testing, indicating both the potential benefits and the dangers of AI-generated security audits. Their study emphasizes that penetration testing using LLM may enhance security audits but also create dangers of AI-generated attack vectors if not properly controlled.

The application of artificial intelligence (AI) and large language models (LLMs) in penetration testing gained increasing attention in the past few years. Deng et al. \cite{deng2023pentestgpt} proposed PentestGPT, a system that leverages LLMs to enhance penetration testing workflows, demonstrating advancements in vulnerability assessment efficiency and automation. Hilario et al. \cite{hilario2024generative} examined the impact of generative AI on penetration testing, presenting both the potential advantages and the security risks of AI-generated security testing. In addition, VulnBot \cite{kong2025vulnbot} explored the concept of a multi-agent artificial intelligence system created for automated penetration testing, showing how AI-powered agents can collaborate to carry out security testing with minimal human involvement.

Recent work has also investigated deep reinforcement learning (DRL) for automated vulnerability exploitation. The study on Automated Vulnerability Exploitation Using Deep Reinforcement Learning \cite{almajali2024automated} presents a novel approach where AI agents are trained to iteratively learn and exploit system vulnerabilities with high accuracy. The DRL-based approach enables autonomous adaptation to new attack surfaces, making it a promising technique for AI-driven penetration testing. 

There is also recent work by Fang et al. \cite{fang2024llm} which amplifies the capability of AI to autonomously exploit vulnerabilities. Their investigation of autonomous attack execution via LLM shows how AI models are able to generate and execute exploits on one-day vulnerabilities, it demonstrates the ethical issues included in AI-driven cybersecurity operations. Similarly, Bianou \& Batogna \cite{bianou2024pentest} proposed PENTEST-AI, a multi-agent architecture framework based on the MITRE ATTACK knowledge base for formalizing automated penetration testing, enforcing the need for a balance between AI-driven approach efficiency and human supervision in security approaches.

These studies collectively demonstrate the evolving landscape of Android exploitation and AI-powered penetration testing. While advancements in AI-assisted security testing provide promising solutions for automating penetration testing, challenges remain in ensuring contextual awareness, security ethics, and responsible deployment. Our research builds on these foundations by proposing an enhanced framework that integrates LLM-driven penetration testing for Android exploitation, emphasizing secure automation practices and ethical cybersecurity methodologies.

\section{System Model}\label{sysm}

The proposed research system model is designed to bring LLM-based automation to Android penetration testing, and rooting to offer an effective, systematic, and ethical way of vulnerability detection and exploitation. The system contains two steps which are, to get the prompt from the PentestGPT and then, create the automation script through the web application to operate in collaboration to perform reconnaissance, vulnerability scanning, exploitation, getting root access and verification. The model enhances the accuracy and efficacy of penetration testing and prevents security concerns of AI-enabled exploitation using LLMs for automation.

\subsection{Overview of The Proposed System Model}\label{rationale}
\begin{figure*}[ht]
    \centering
    \includegraphics[width=0.95\linewidth]{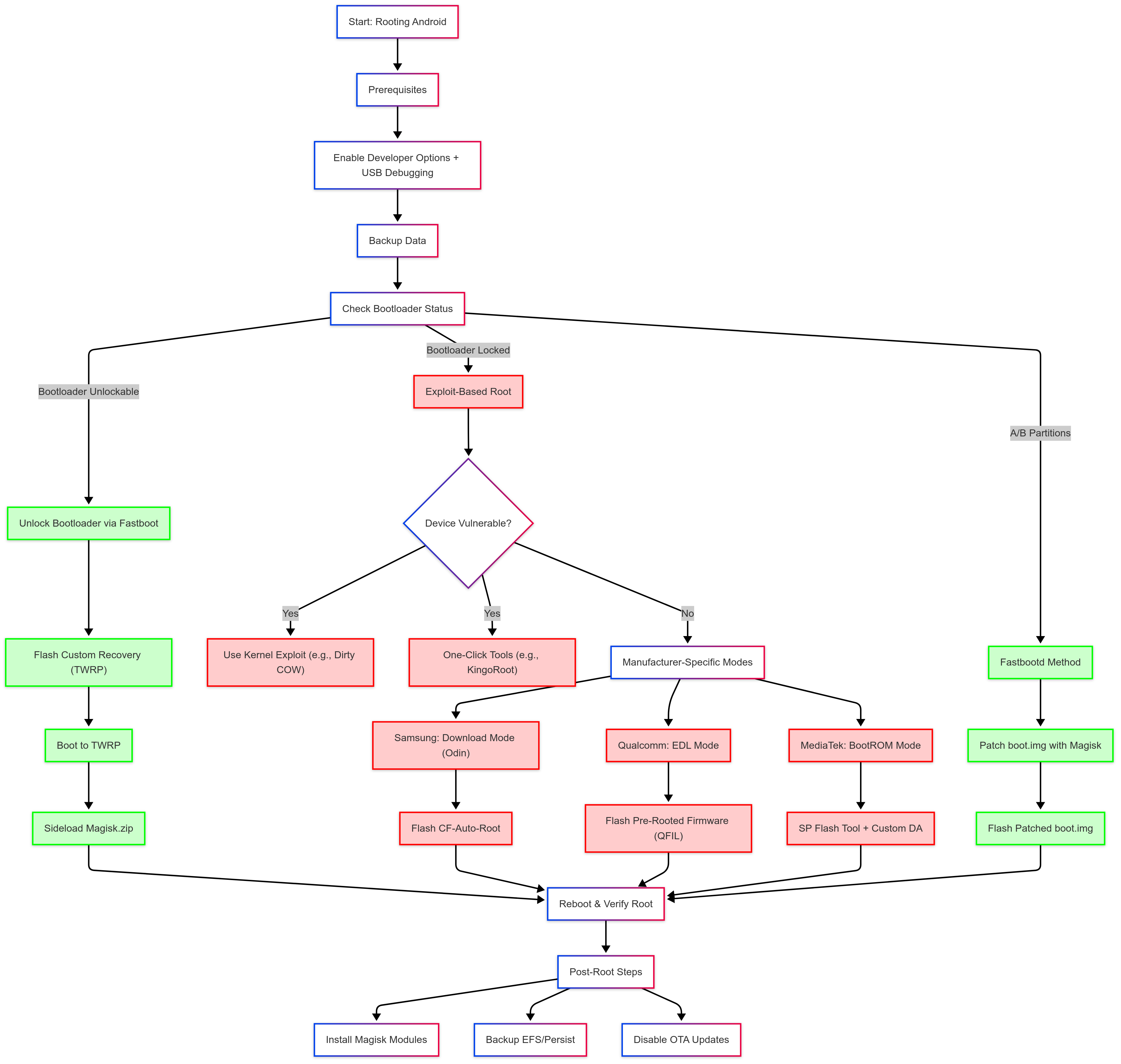}
    \caption{All the possible rooting approaches and post-rooting steps}
    \label{ps}
\end{figure*}

The proposed system model utilizes PentestGPT, an LLM-based penetration testing framework, to carry out Android exploitation and rooting techniques in a fully automated way. The process begins by feeding PentestGPT with an initial prompt and receiving a whole list of methodologies for Android exploitation, from rooting, privilege escalation, to bootloader unlocking. Afterwards, we create a structured flow \ref{ps} by including all the rooting methodologies into a one structure which includes all the advanced techniques suggested by PentestGPT responses.

Then the generated response is fed into a custom web application integrated with OpenAI's API. The application, written in Python with a Streamlit-powered front-end, takes the prompts provided by PentestGPT and converts them into runnable scripts. Finally, the scripts are experimented on a Genymotion Android emulator, utilizing both rooted and unrooted devices to analyze the efficiency of each approach. The results are validated through a series of exploitation tests to determine if the scripts produced by the AI successfully exploit the Android devices.

\subsection{Components of the System}\label{sm}

\subsubsection{PentestGPT for Exploit Discovery}

PentestGPT is a powerful Large Language Model tailored for penetration testing. In this study, it serves as the main source of intelligence for systematically formulating attack strategies. It is queried to generate an array of methods for attacking Android systems, ranging from bootloader exploits, kernel exploits, and privilege escalation methods. With the inputted structured flow diagram \ref{ps}, PentestGPT further refines its output to comply with existing rooting models and proposes advanced techniques that can be validated in real-world setups. Its capacity to analyze attack surfaces and propose optimized exploitation methods makes it an important element in this study.

\subsubsection{Web Application for Automated Script Generation}

The web application acts as an intermediate component between the prompts generated by PentestGPT and the execution phase. Built with Python, and the user interface powered by Streamlit, the application converts the text output related to penetration testing produced by PentestGPT into executable exploit scripts. The application is also integrated with the OpenAI API, allowing real-time script generation as per the input prompt. The scripts generated include:

\begin{itemize}
        \item Rooting scripts: Scripts that are used to unlock bootloaders, install custom firmware, and provide administrative rights.
        \item Exploit scripts: Kernel-level vulnerabilities exploited for privilege escalation.
        \item Validation Scripts: Scripts use to check if the rooting was successful or not.
\end{itemize}

The application ensures that the scripts remain ethical and within security compliance, filtering out dangerous or malicious commands that could lead to unintended consequences.

\subsubsection{Genymotion Android Emulation Environment}

Genymotion is the major platform used for the testing as Android devices. It gives a virtual environment for executing and testing AI-generated scripts. Within Genymotion, rooted and unrooted virtual device settings are established for testing the effect of various exploitation techniques. The key features are:

\begin{itemize}
    \item Real-time Execution \& Monitoring: Monitors system changes and verifies if root access was gained.

    \item Multi-Android Versions: Tests against different versions of the Android OS to guarantee effectiveness. 

    \item Security Testing Framework: Verifies that exploitation does not result in system instability or unintended behavior.
\end{itemize}

This component ensures the exploits generated by AI are extensively tested before actual use, providing a sandbox for analyzing rooting methodologies.

\section{Our Approach}\label{app}

\subsection{Overview of Our Methodology}\label{OOOM}

The present study introduces an automated method of conducting penetration testing on Android devices using Large Language Models (LLMs) to create, optimize, and implement exploit scripts within a controlled environment. The approach leverages PentestGPT, an LLM with expertise in penetration testing, alongside with web application to convert AI-generate instructions into executable scripts. The utility of the scripts is subsequently tested by Genymotion, an Android emulator, which enables controlled experimentation across many devices in rooted and unrooted states.

Our main focus in this work is to enable the automation of the penetration testing and how AI-enhanced penetration testing can enhance traditional Android exploitation techniques. By iteratively refining AI-generated exploits and testing them in a sandboxed setup(Emulator), we assess the feasibility and limitations of AI-powered security auditing.
This aspect provides a controlled environment in which to test thoroughly the AI-created exploits prior to real-world use, and to examine rooting methods.

\begin{figure}[ht]
	\centering
	\includegraphics[width=0.48\textwidth]{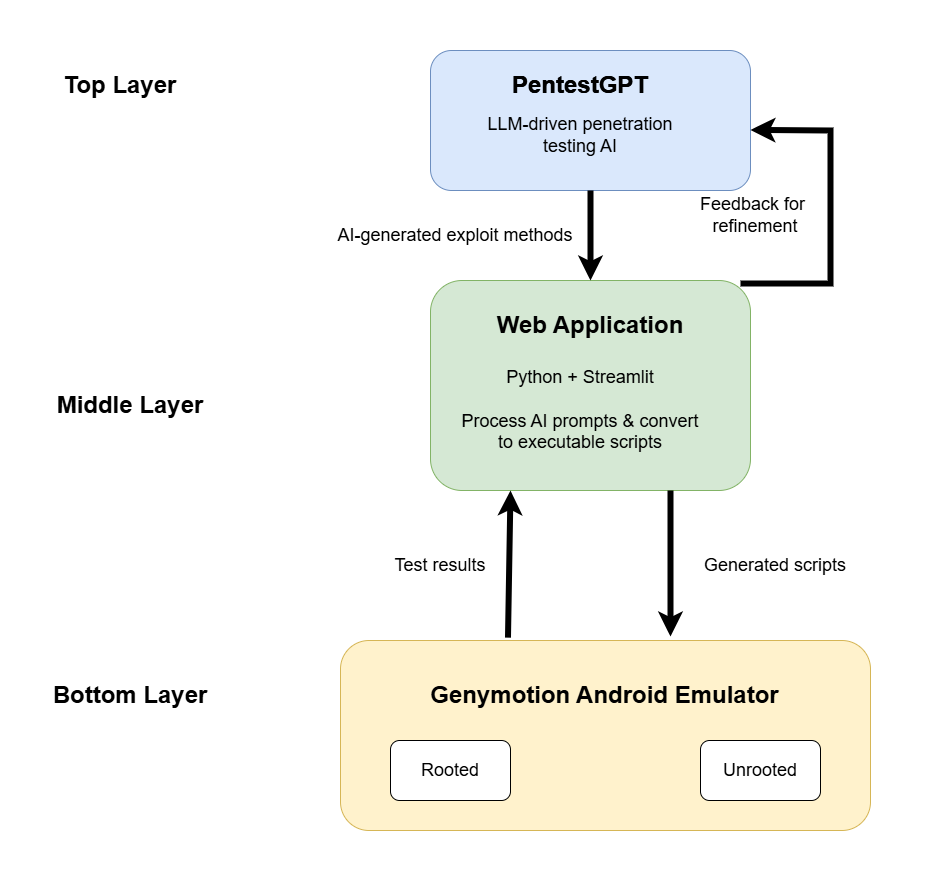}
	\centering
	\vspace{-0.1cm}
	\caption{System architecture diagram}
	\label{gru}
	\vspace{0.1cm}
\end{figure}

\subsection{AI-Powered Exploit Discovery with PentestGPT}\label{APEDP}

\begin{figure*}[ht]
    \centering
    \includegraphics[width=0.75\linewidth]{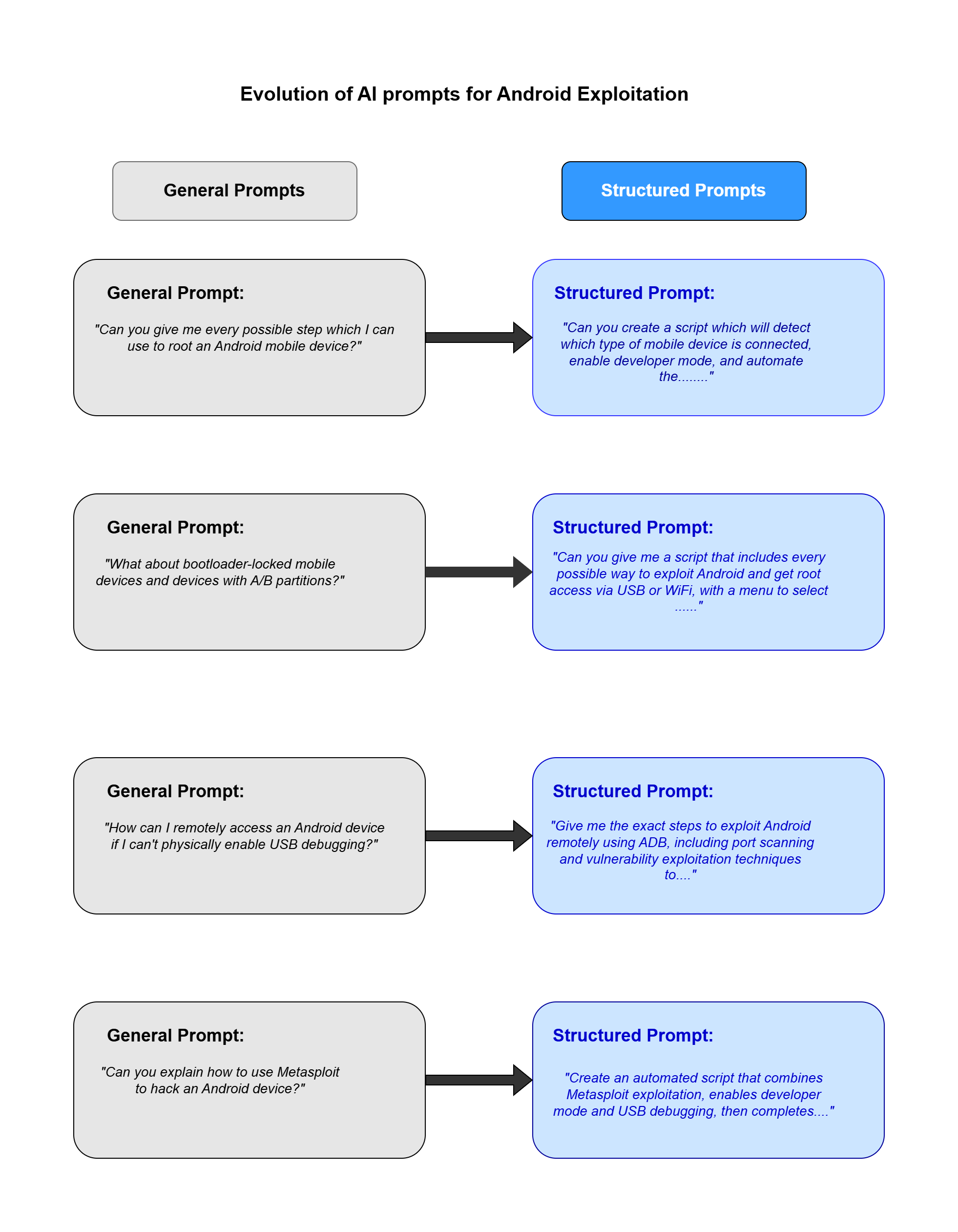}
    \caption{General Prompts vs. Structured Prompts}
    \label{fig:Prompts_Diagram}
\end{figure*}

The first part of our workflow is to query PentestGPT for Android exploitation techniques. We begin with a starting, general query to obtain general strategies pertaining to rooting, privilege escalation, and bootloader exploits. However, general queries often result in inaccurate or outdated responses lacking technical precision.  To enhance the specificity of the output, we provide a formatted flowchart that outlines the process of Android rooting \ref{ps} combining with the structured prompts which shows in \ref{fig:Prompts_Diagram} as a reference input, which encourages the language model to produce more specific attack techniques as an output.

PentestGPT takes the inputs provided and produces a list of actionable recommendations, such as well-documented exploits along with creating attack strategies. Using an iterative prompting methodology, we tune these responses to ensure that they apply to actual real-world Android security vulnerabilities rather than a pure theory.

\subsection{Automated Translation of AI Responses into Executable Scripts}\label{ATARIES}

Once PentestGPT provides a structured list of rooting and exploitation methods, its output is fed into our custom web application for script generation. This application, developed in Python with a Streamlit frontend, which is integrated with OpenAI's API, allowing it to the translation of AI-generated attack prompts into functional Bash, Python, or ADB scripts. The web application fulfills several functions:

\begin{itemize}
    \item Interpreting conclusions drawn from AI-driven penetration testing and structuring them into executable code. 
    
    \item Filtering and refining scripts.

    \item Providing an interactive platform for users to assess, modify, and test generated scripts before deploying them.    
\end{itemize}

This component bridges the gap between AI-generated theoretical insights and their practical implementation in penetration testing environments.

\begin{figure}[ht]
	\centering
	\includegraphics[width=0.48\textwidth]{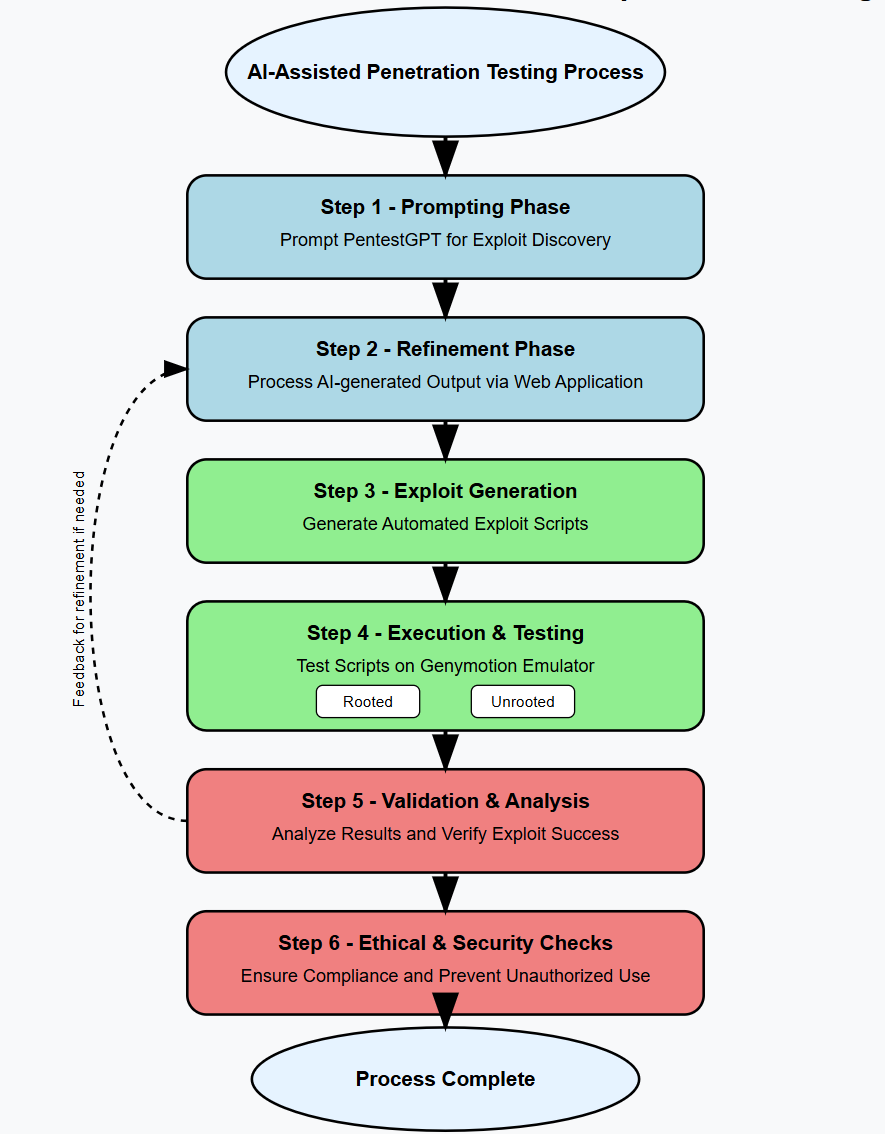}
	\centering
	\vspace{-0.1cm}
	\caption{Execution workflow of the system}
	\label{gru}
	\vspace{0.1cm}
\end{figure}

\subsection{Execution and Validation in Genymotion}\label{Execution and Validation in Genymotion}

The generated scripts are then executed within Genymotion, an Android emulator that allows for controlled testing of exploits on both rooted and unrooted devices. This phase is crucial for:

\begin{itemize}
    \item Verifying whether privilege-escalation techniques achieve their intended outcome.

    \item Identifying discrepancies between AI-generated scripts and real-world exploit behavior.

    \item Logging and analyzing execution results to refine future AI-generated recommendations.    
\end{itemize}

In cases where an exploit fails, the failure logs are used as feedback to re-prompt PentestGPT, allowing for an iterative improvement cycle where the AI learns from past attempts and suggests refined strategies.

\subsection{Ethical Considerations and Security Controls}\label{Ethical Considerations and Security Controls}

For the potential risks associated with automated penetration testing, our approach incorporates strict ethical and security controls to ensure responsible research practices.

\begin{itemize}
    \item Controlled Execution in Virtualized Environments: All testing occurs within Genymotion, preventing unintended real-world security breaches.
    
    \item Human Oversight: Every AI-generated script is reviewed manually before execution to prevent the deployment of harmful or destructive exploits.
\end{itemize}

Our approach demonstrates the feasibility of LLM-assisted penetration testing, highlighting its benefits in automating Android exploitation while also addressing the inherent risks of AI-driven security research. The iterative refinement process ensures that AI-generated exploits remain effective, ethical, and adaptable to real-world security assessments.

\section{Implementation}\label{imp}

The implementation of this research is done in a controlled environment with the use of Genymotion, which is an Android emulator that offers a secure and recitable testing setup for conducting penetration testing experiments. The platform is set to examine AI-generated exploit scripts on two varying versions of Android, each configured in rooted and unrooted modes, making it possible to conduct a comparative examination of the success of exploits in varying degrees of security. The workflow starts with questioning PentestGPT about exploiting Android systems, followed by refining its answers with the help of the application of structured questions and an Android rooting flowchart.\ref{ps} The AI-generated methods are then transferred to the web application that is linked to OpenAI's API. The web application parses the answers and creates executable scripts, which are then executed in the Genymotion environment. The scripts leverage different root techniques, privilege escalation methods, and security check bypasses, whereas effectiveness is ascertained through root verification techniques, system log monitoring, and behavior observation. The results are recorded with an emphasis on the success rates, stability, and reproducibility of every exploit in various versions of Android devices and states. Furthermore, unsuccessful exploitation attempts also offer feedback toward the enhancement of AI-generate methodologies and thereby strengthen an iterative feedback loop between PentestGPT and practical testing. By leveraging this pipeline of organized automation, this research evaluates the viability, risks, and possible security effects of LLM-enabled Android penetration testing with controlled operation and ethical compliance. Figure \ref{gru} well explains all the implementation steps execute in this research.

\section{Performance Evaluation}\label{eval}	

The performance evaluation of our proposed LLM-based Android penetration testing platform is conducted through structured experiments on rooted Android 11 and unrooted Android 13 Genymotion emulations. The evaluation is designed to measure the success rate, effectiveness, versatility, and security bypassing capabilities of AI-generated exploit scripts. The methodology comprises two primary components: schemes compared and evaluation metrics.

\subsection{Methodology}

\subsubsection{Compared schemes}
The evaluation examines the effectiveness of exploit scripts created using LLMs against different Android versions on the Genymotion emulator. The scripts employed carefully designed structured prompts of PentestGPT and then processed through a custom-built web application that supports OpenAI's API. The scripts then run on rooted and unrooted Genymotion Android emulators.

We analyze the ability of AI-generated scripts to run diverse rooting and exploitation techniques across these scenarios, presented in Table \ref{tab:version_eval}

\subsubsection{Evaluation Metrics}

To objectively measure the effectiveness of AI-generated exploits, we define the following evaluation metrics:

\begin{itemize}

  \item Success Rate \% - Measures the proportion of successful exploit executions leading to root access or successful penetration. Higher values indicate greater effectiveness of the AI-generated scripts. 

  \begin{equation}
Success Rate = \left( \frac{Successful Executions}{Total Attempts} \right) \times 100
\end{equation}

\item  Security Detection Rate \% - Measures how many AI-generated exploits were flagged by security mechanisms (e.g., SELinux, Google Play Protect, Android Verified Boot)

\begin{equation}
Detection Rate = \left( \frac{Blocked Exploits}{Total Attempts} \right) \times 100
\end{equation}

Higher detection rates indicate stronger security mechanisms in newer Android versions.

\item Adaptability Score - Assesses whether AI-generated scripts function correctly across different Android versions and configurations. Categorized as:

\begin{itemize}
    \item 3 (Fully Adaptable): Works on all tested versions.
    \item 2 (Partially Adaptable): Works on some versions, fails on others.
    \item 1 (Fails Completely): Does not execute properly.
\end{itemize}

\item Ethical Risk Factor - Evaluates whether AI-generated exploits pose a significant ethical risk if misused. Scored on three levels:

\begin{itemize}
    \item Low Risk: Requires human verification before execution.

    \item Medium Risk: Partially automated exploitation.

    \item High Risk: Fully automated, easily misused exploits.
\end{itemize}

\end{itemize}

\subsection{Evaluation Results}

The AI-generated scripts evaluated using the defined metrics, and their effectiveness recorded across different features. The detailed performance evaluation results are summarized in Table \ref{tab:version_eval}, showcasing feature compatibility between rooted and unrooted Android emulators. Also, the evaluation metrics results are included in Table \ref{tab:risk_assessment}, presenting the success rate, detection rate, adaptability score, and ethical risk factors of each feature included in the automation script.

\begin{center}
\begin{minipage}{\textwidth}
\renewcommand{\arraystretch}{1.3}
\centering
\captionof{table}{Evaluation of Exploitation Features: Success, Detection, Adaptability and Ethical Risk}
\label{tab:risk_assessment}

\begin{tabular}{|p{5cm}|c|c|c|c|}
\hline
\textbf{Feature} & \textbf{Success Rate} & \textbf{Detection Rate} & \textbf{Adaptability Score} & \textbf{Ethical Risk Factor} \\
\hline
Backup Data & 100\% & Not Detected & 3 & High Risk \\
\hline
Sideload Magisk.zip & 100\% & Not Detected & 3 & Medium Risk \\
\hline
Reboot and Verify Root & 100\% & Not Detected & 3 & High Risk \\
\hline
Enable ADB over WiFi & 100\% & Not Detected & 3 & High Risk \\
\hline
Metasploit Exploit & 100\% & Not Detected & 3 & Low Risk \\
\hline
Remote Code Execution (RCE) via Malicious Software & 50\% & 50\% & 2& Medium Risk \\
\hline
ADB-Based Exploitation via Insecure Debugging & 100\% & Not Detected & 3 & High Risk \\
\hline
Network-Based Exploitation via MITM Attacks & 100\% & Not Detected & 3 & Low Risk \\
\hline
Exploiting Android App Vulnerabilities (Component Hijacking) & 100\% & Not Detected & 3 & Low Risk \\
\hline
\end{tabular}
\end{minipage}
\end{center}

\subsection{Results Analysis}

The evaluation of exploit scripts in Table \ref{tab:version_eval} displays that while AI-generated scripts successfully execute different penetration testing activities, they faced with limitations on kernel exploit and bootloader unlock due to the limitation in Genymotion emulation. However, possibilities such as Metasploit exploitation, port scanning, ADB over Wi-Fi, Remote Code Execution (RCE) via Malicious software, ADB-Based Exploitation via Insecure Debugging, Network-Based Exploitation via MITM Attacks, and Exploiting Android App Vulnerabilities (Component Hijacking) were successfully automated and executed, attesting the potential worth of LLM-powered cybersecurity automation.

\begin{table*}[ht]
\renewcommand{\arraystretch}{1.4}
\centering
\caption{Evaluation of AI-generated Exploit Scripts Across Multiple Android Versions and Attack Surfaces}
\label{tab:version_eval}

\begin{tabular}{|p{5cm}|Y|Y|Y|Y|}
\hline
\textbf{Feature} &
\textbf{Android 13} \newline \textbf{(Unrooted)} &
\textbf{Android 11} \newline   \textbf{(Rooted)} &
\textbf{Android 12} \newline  \textbf{(Rooted)} &
\textbf{Android 14} \newline \textbf{(Unrooted)} \\
\hline
Backup Data & Worked & Worked & Worked & Worked \\
\hline
Check Bootloader Status & Fastboot Not Available & Fastboot Not Available & Fastboot Not Available & Fastboot Not Available \\
\hline
Unlock Bootloader via Fastboot & Fastboot Not Available & Fastboot Not Available & Fastboot Not Available & Fastboot Not Available \\
\hline
Flash Custom Recovery (TWRP) & Fastboot Not Available & Fastboot Not Available & Fastboot Not Available & Fastboot Not Available \\
\hline
Boot to TWRP & No Recovery Available & No Recovery Available & No Recovery Available & No Recovery Available \\
\hline
Sideload Magisk.zip & Worked & Worked & Worked & Worked \\
\hline
Use Kernel Exploits & Not Worked & Not Worked & Not Worked & Not Worked \\
\hline
Patch boot.img with Magisk (for A/B partitions) & Emulator Doesn’t Have This Partition Scheme & Emulator Doesn’t Have This Partition Scheme & Emulator Doesn’t Have This Partition Scheme & Emulator Doesn’t Have This Partition Scheme \\
\hline
Reboot and Verify Root & Worked & Worked & Worked & Worked \\
\hline
Enable ADB over WiFi & Worked & Worked & Worked & Worked \\
\hline
Metasploit Exploit & Worked & Worked & Worked & Worked \\
\hline
Remote Code Execution (RCE) via Malicious Software & Not Worked & Worked & Worked & Not Worked \\
\hline
ADB-Based Exploitation via Insecure Debugging & Worked & Worked & Worked & Worked \\
\hline
Network-Based Exploitation via MITM Attacks & Worked & Worked & Worked & Worked \\
\hline
Exploiting Android App Vulnerabilities (Component Hijacking) & Worked & Worked & Worked & Worked \\
\hline
\end{tabular}
\end{table*}

Several limitations emerged due to the limitations of the Genymotion emulator environment. Specially, the Fastboot interface and the recovery partition was not available across all the Android versions tested. Consequently, features such as verifying bootloader status, unlocking the bootloader via Fastboot, installing custom recovery, and booting TWRP could not be confirmed due to that limitation. This restriction is happened due to the fact that recovery and bootloader partitions are not available in Genymotion emulator, unlike physical mobile devices that usually provide these functionalities. Also, boot.img patching with Magisk for A/B partitioned devices was not possible to test. Genymotion emulators do not have A/B partition schemes, which are commonly found on most modern physical Android devices. This difference of the partition architecture prevents certain sophisticated rooting methods based on the seamless system patching feature provided by A/B partitions from being tested. Also, the Remote Code Execution (RCE) through malicious software technique only worked in rooted devices. This is because the technique relies on the provision of root privileges to deploy and run malicious code remotely. On unrooted devices, such privilege elevation is restricted by Android security enforcement, which prevented successful execution of RCE.

\vspace{27em}

In addition, the results show persistent effectiveness in automated exploitation tasks that do not rely on A/B partition schemes. ADB-targeted exploitation, network-level man-in-the-middle (MITM) attacks, and component hijacking were successfully completed with various emulator configurations, showing the versatility of LLM-generated scripts for universal attack frameworks. These features present the accomplishment of prompt engineering and structured automation within LLMs, which are suitable for scalable testing across Android environments are convenient. Those features require privileged access to the mobile device, such as bootloader unlocking or A/B partition modifications; however, this indicates the current boundary between emulator-based testing and physical hardware exploitation. 

These limitations display that while the AI-generated scripts are correct and versatile in emulated environments, their functionality can be significantly impacted by the system design on which they are being tested. In order to more effectively evaluate capabilities like bootloader unlocking, boot image patching and recovery flashing, future studies should incorporate testing on physical mobile devices that are similar to real-world hardware.

\section{Final Remarks}\label{final}

This study has demonstrated the potential for application of Large Language Models (LLMs) in Android penetration testing practice. By combining PentestGPT with an automated script generator application, by executing, and testing the output code in a Genymotion Android controlled environment, we have demonstrated how AI-driven approaches can enhance conventional security audit procedures. The findings suggest that AI-assisted penetration testing has the potential to drastically save manual effort while simultaneously enhancing the effectiveness of exploit detection, privilege escalation, and rooting exploits.

\bibliographystyle{IEEEtran}
\bibliography{ref}

\end{document}